\documentclass{emulateapj}

\shorttitle{The Nature Of M31 Halo Substructure}
\shortauthors{J.C. Richardson~et al.}

\begin{document}

\slugcomment{Accepted to AJ}

\title{THE NATURE AND ORIGIN OF SUBSTRUCTURE IN THE OUTSKIRTS OF M31
\\I: SURVEYING THE STELLAR CONTENT WITH HST/ACS\footnotemark[1]}
\footnotetext[1]{Based on observations made with the NASA/ESA Hubble
Space Telescope, obtained at the Space Telescope Science Institute,
which is operated by the Association of Universities for Research in
Astronomy, Inc., under NASA contract NAS 5-26555.}

\author{
  J.\,C.\,Richardson\altaffilmark{2},
  A.\,M.\,N.\,Ferguson\altaffilmark{2},
  R.\,A.\,Johnson\altaffilmark{3},
  M.\,J.\,Irwin\altaffilmark{4},
  N.\,R.\,Tanvir\altaffilmark{5},
  D.\,C.\,Faria\altaffilmark{4},
  R.\,A.\,Ibata\altaffilmark{6},
  K.\,V.\,Johnston\altaffilmark{7},
  G.\,F.\,Lewis\altaffilmark{8}
}

\altaffiltext{2}{Institute for Astronomy, University of Edinburgh,
  Royal Observatory, Blackford Hill, Edinburgh EH9 3HJ, UK}\email{jcr@roe.ac.uk}
  \altaffiltext{3}{Department of Astrophysics, University of Oxford,
  Keble Road, Oxford, OX1 3RH, UK}  \altaffiltext{4}{Institute of
  Astronomy, University of Cambridge, Madingley Road, Cambridge, CB3
  0HA, UK}
  \altaffiltext{5}{Department of Physics and Astronomy, University of
  Leicester, LE1 7RH, UK}
  \altaffiltext{6}{Observatoire de Strasbourg, 11, rue de
  l'Universit\'{e}, F-67000 Strasbourg, France}
 \altaffiltext{7}{Department of
  Astronomy, Wesleyan University, Middletown, CT 06459, USA}
  \altaffiltext{8}{Institute of Astronomy, School of Physics, A29,
  University of Sydney, NSW 2006, Australia}


\begin{abstract} 
  We present the largest and most detailed survey to date of the
stellar populations in the outskirts of M31 based on the homogeneous
analysis of 14 deep HST/ACS pointings spanning the range $11.5$~kpc
$\lesssim$ R$_{proj} \lesssim 45$~kpc. Many of these pointings sample
coherent substructure discovered in the course of the INT/WFC imaging
survey of M31 while others sample more diffuse structure in the
extended disk.  We conduct a quantitative comparison of the resolved
stellar populations in these fields and identify several striking
trends. The color-magnitude diagrams (CMDs), which reach $\gtrsim 3$
magnitudes below the red clump with high completeness, can be
classified into two main categories based on their
morphologies. `Stream-like' fields, so named for their similarity to
the CMD of the giant stellar stream, are characterized by a red clump
that slants bluewards at fainter magnitudes and an extended horizontal
branch. They show no evidence for young populations.  On the other
hand, `disk-like' fields exhibit rounder red clumps with significant
luminosity width, lack an obvious horizontal branch and show evidence
for recent star formation ($\sim 0.25-2~{\rm Gyr}$ ago).  We compare
the spatial and line-of-sight distribution of stream-like fields with
a recent simulation of the giant stream progenitor orbit and find an
excellent agreement. These fields are found across much of the inner
halo of M31, and attest to the high degree of pollution caused by this
event.  Disk-like material resides in the extended disk structure of
M31 and is detected here up to ${\rm R_{proj}} \sim 44 ~{\rm kpc}$;
the uniform populations in these fields, including the ubiquitous
presence of young populations, and the strong rotation reported
elsewhere are most consistent with a scenario in which this structure
has formed through heating and disruption of the existing thin disk,
perhaps due to the impact of the giant stream progenitor.  Our
comparative analysis sheds new light on the likely composition of two
of the ultra-deep pointings formerly presented as pure outer disk and
pure halo in the literature.
\end{abstract}



\section{Introduction}
\label{sec:intro}

A key goal of modern astrophysics is to understand the formation
history of galaxies like our own Milky Way. The favored paradigm of
hierarchical assembly within a cold dark matter ($\Lambda$CDM)
dominated universe predicts that spheroidal galaxy components form
through a repetitive process of galaxy mergers and the accretion of
smaller subsystems while disks arise from the smooth accretion of gas
\citep[e.g.,][]{WhiteFrenk91,SteinmetzNavarro02}).  Recently, much
ground has been gained in understanding the nature of stellar halos
built up by tidal stripping of accreted sub-halos through N-body plus
semi-analytical models \citep{BJ05, Font06} and through numerical
simulations \citep{HelmiWhite99}.

Assuming the accreted systems possess a stellar component, the most
visible evidence of this hierarchical formation is expected in the
form of tidal tails and partially digested satellites in the extended
halos of galaxies \citep[e.g.][]{BJ05}. Theoretical work on the tidal
destruction of satellites by their massive hosts has shown that the
mixing time of the debris depends on the satellites mass and
orbit. This timescale can be relatively short for debris in the inner
galaxy but many gigayears for the outer galaxy \citep{JHB96}.
Searches for spatially coherent tidal features in the extended halos
of galaxies thus provide a means to probe $\Lambda$CDM predictions on
scales of individual galaxies. Once tidal features are found, detailed
follow-up studies of their stellar content and kinematics provide
important constraints on the nature and number of objects which have
merged. The Milky Way hosts at least one major tidal stream,
originating from the Sagittarius dwarf \citep{Majewski03}, with many
others suggested \citep[e.g.,][]{Grillmair06, Belokurov07}.

\begin{figure}[h!]
\includegraphics[width=11.0cm,angle=-90]{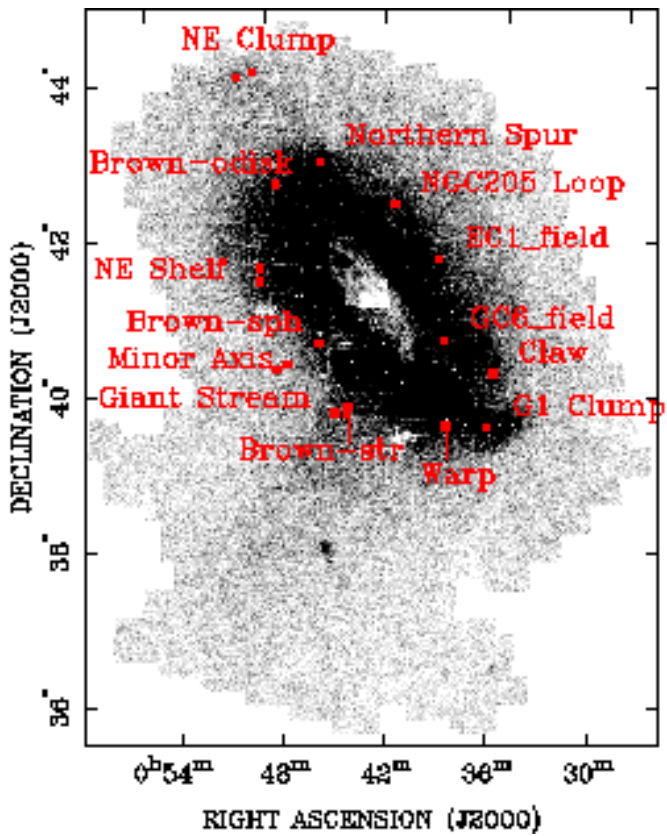}
  \caption{Map showing the distribution of RGB stars around M31 from
  the INT/WFC survey \citep{Irwin05}. The image spans $\sim 95~{\rm
  kpc} \times 125~{\rm kpc}$. The locations of our HST/ACS pointings
  are overlaid in red, each covering $0.8~{\rm kpc} \times 0.8~{\rm
  kpc}$ at the distance of M31.}
  \label{fig:map}
\end{figure}

It can be argued, however, that the best current laboratory for
studying faint stellar substructure around galaxies is provided by our
nearest giant neighbor, M31. Our external view of this system removes
complications due to line-of-sight and extinction/crowding effects
that plague Milky Way studies and as a result the structures observed
in M31 are considerably easier to interpret. M31 has the been the
subject of several deep wide-field ground-based surveys in the last
decade and abundant low-surface brightness stellar substructure has
been found.  These surveys have the depth to resolve individual red
giant branch (RGB) stars in the halo of M31 and the analysis of their
large-scale spatial density distribution has allowed unprecedented
surface brightness levels to be reached ($\Sigma_V \gtrsim 30$
magnitudes per square arcsecond).  The Isaac Newton Telescope
Wide-Field Camera (INT/WFC) panoramic survey was the first of these
surveys to be conducted and imaged the outer disk and inner halo of
M31 out to a radius of $\sim 55 ~{\rm kpc}$ revealing copious
substructure and tantalizing evidence for metallicity variations
(Figure 1, see also \cite{Ibata01a, F02} and \cite{Irwin05}). In the
southern quadrant, a giant stream (hereafter GS) was discovered
falling into the center of M31 from behind \citep{Ibata01a, McC03,
Ibata04}.  Prominent stellar overdensities were also discovered along
both major axes, beyond the extent of the bright disk.  More recently,
the entire southern quadrant of M31 has been imaged out to a radius of
$\sim 150 ~{\rm kpc}$, with an extension to the galaxy M33 at $\sim
200 ~{\rm kpc}$, using Megacam on the Canada France Hawaii Telescope
(CFHT) \citep{Martin06, Ibata07}. This survey has revealed an
additional complex system of even fainter tidal streams in the far
outer halo, including a series of azimuthal streams along the minor
axis out to $\sim120 ~{\rm kpc}$. The Megacam survey has probed the GS
out to a projected radius of $\sim 100 ~{\rm kpc}$ and uncovered
evidence for internal stellar populations variations.  Current
simulations suggest it can be explained as the trailing debris arm of
a disrupted dwarf galaxy of mass $\sim 10^9 ~{\rm M}_{\sun}$ whose
leading arm may have wrapped around the inner galaxy at least twice
(\cite{MoriRich08, Fardal07} and references therein). The highly
radial orbit of the stream suggests it passes very close to the center
of M31 \citep{Ibata04} and hence almost certainly interacts with the
disk.  This may represent one of the most significant accretion events
since the initial formation of M31, providing us a fortuitous glimpse
of ongoing mass assembly in the z = 0 universe. The presence of the GS
has raised many questions: What role has this particular event played
in the late formation history of M31? How much of the other inner halo
debris has come from the stream?  What damage has the stream
progenitor done to the M31 disk?  To what extent have other recent
satellite accretions contributed to the growth of the halo?

As a means to address these questions, we have been using the Advanced
Camera for Surveys (ACS) on board the Hubble Space Telescope (HST) to
develop a better understanding of the origin and nature of the stellar
substructure observed around M31. \cite{F05} presented the first
detailed color magnitude diagrams (CMDs) of six regions of prominent
inner halo substructure, reaching several magnitudes below the red
clump (RC). Analysis revealed distinct variations in the stellar
populations throughout many of the regions studied, consistent with
the substructure having more than one origin. Two of the regions
studied, the NE Shelf and the GS, were shown to have remarkably
similar populations but different line-of-sight distances, consistent
with the shelf being a forward wrap of the stream \citep{F05}. A
detailed examination of the stellar populations in one of the
prominent major axis overdensities, the G1 Clump, was presented by
\cite{Faria07}.  This field, which lies at $\sim 30 ~{\rm kpc}$, was
shown to have experienced continuous yet declining star formation
during the last 10 Gyr and have a relatively high metallicity of
[M/H]$=-0.4$ dex. \cite{Faria07} argued that these properties are more
consistent with those of the M31 outer disk than a low mass accreted
dwarf and suggested this piece of substructure could have been
recently torn off from the main disk. \cite{Ibata05} have studied the
kinematics of many of these features, finding an overall pattern of
significant rotation.  They hypothesize that this extended and highly
structured rotating structure has been formed by accretion.

\begin{deluxetable*}{c c c c c c c c}[ht!]
  \tablewidth{17.5cm} \tablecaption{Observational
    Information\label{tab:obs}} \tablehead{\colhead{Field} &
    \colhead{Proposal ID} & \colhead{P.I.} & \colhead{R.A. (J2000)} & \colhead{Dec
    (J2000)} & \colhead{Date} & \colhead{$t_{F606W}$ (s)
    \tablenotemark{a}} & \colhead{$t_{F814W}$ (s) \tablenotemark{b}}  }
    \startdata

    NGC205 Loop & GO9458 & A. Ferguson & 00:41:11.6 &  42:29:43.1 &  2003-02-25 & 2475 & 5290 \\
    Minor Axis  & GO9458 & A. Ferguson & 00:48:08.4 &  40:25:30.0 &  2003-08-09 & 2400& 5100  \\  
                & GO9458 & A. Ferguson & 00:48:47.8 &  40:20:44.6 &  2003-08-04/05 &  2400 & 5100 \\  
    NE Shelf    & GO9458 & A. Ferguson & 00:49:59.4 &  41:28:55.5 &  2003-07-31 &  2475 & 5290 \\
                & GO9458 & A. Ferguson & 00:50:05.7 &  41:39:21.4 &  2003-08-07 &  2400 & 5100 \\
    G Stream    & GO9458 & A. Ferguson & 00:44:15.5 &  39:53:30.0 &  2003-01-08 & 2430 & 5150 \\
                & GO9458 & A. Ferguson & 00:45:05.0 &  39:48:00.0 &  2002-10-17 & 2430 & 5150\\
    N Spur      & GO9458 & A. Ferguson & 00:46:10.0 &  43:02:00.0  & 2003-07-08 & 2475 & 5290 \\
    G1 Clump    & GO9458 & A. Ferguson & 00:35:28.0 &  39:36:19.1  & 2003-01-17 & 2430 & 5150 \\
    Warp        & GO9458 & A. Ferguson & 00:38:05.1 &  39:37:54.9  & 2003-06-10/11 & 2620 & 5240 \\
    Claw        & GO10128 & A. Ferguson & 00:35:00.3 &  40:17:37.3 & 2005-01-06 & 2469 & 5210 \\
    NE Clump    & GO10128 & A. Ferguson & 00:51:56.7 &  44:06:38.5 & 2004-10-12 & 2469 & 5210 \\
                & GO10128 & A. Ferguson & 00:50:55.2 &  44:10:00.4 & 2004-10-09 & 2469 & 5210 \\
    EC1\_field   & GO10394 & N. Tanvir   & 00:38:19.5 &  41:47:15.4 & 2005-07-10 & 1809 & 3000 \\
    GC6\_field  & GO10394 & N. Tanvir   & 00:38:04.6 &  40:44:39.8 & 2005-07-23 & 1809 & 3000 \\
    Brown-stream     & GO10265 & T. Brown    & 00:44:18.0 &  39:47:36.0 & 2004-09-03/05 & 2460 & 5200  \\
    Brown-spheroid     & GO9453  & T. Brown    & 00:46:08.1 &  40:42:36.4 & 2002-12-02/03 & 2460 & 5200 \\
    Brown-odisk     & GO10265 & T. Brown    & 00:49:08.5 &  42:44:57.0 & 2004-12-20/22 & 2460 & 5200 \\
    \enddata
\tablenotetext{a}{Total exposure time in F606W filter. Note that the Warp and Brown fields analyzed here do not reach their full depths.}
\tablenotetext{b}{Total exposure time in F814W filter. Note that the Warp and Brown fields analyzed here do not reach their full depths.}

\end{deluxetable*}

In parallel to our efforts to understand the large-scale structure and
stellar content of M31's outskirts, a number of ultra-deep pencil beam
pointings have been obtained with HST/ACS \citep{Brown03, Brown06a,
Brown06b, Brown07}.  With upwards of 32 orbits per pointing, these
observations have been sufficiently deep to enable stars to be
resolved to and below the oldest main-sequence turn-off, allowing the
complete reconstruction of the star formation history (SFH) in
selected locations. \cite{Brown06a} present results from inner halo
fields they associate with the `Tidal Stream', `Outer Disk' and
`Spheroid'. In all cases, the fields were shown to have experienced an
extended SFH, and differences between them were quantified. Their
Spheroid field, which lies $\sim$ 11~${\rm kpc}$ along the southern
minor axis, was shown to be metal rich ($<$[Fe/H]$>$ $ \gtrsim$ -0.6
dex) and with a substantial intermediate age component. Interestingly,
their Tidal Stream field, which lies very close to the GS field
observed by \cite{F05}, showed almost identical populations (but being
$\sim 1 ~{\rm Gyr}$ younger).  Finally, the Outer Disk field was best
fit by populations of still higher metallicities and ages in the range
4-8 Gyr.  The interpretation of these results has since been
complicated by the suggestion that much of the inner halo, and in
particular the southern minor axis, is `contaminated' by tidal debris
torn off from the GS progenitor \citep[e.g.,][]{Gilbert07,
Ibata07}. Indeed, in this paper we highlight the global extent of GS
progenitor debris across the face of M31.

The halo of M31 is clearly highly complex making it difficult to draw
firm conclusions about the galaxy's assembly history from the study of
only a few small fields. For example, it has recently been suggested
that the studies claiming the stellar halo of M31 is markedly
different from that of the Milky Way may have drawn inaccurate
conclusions due to comparing kinematically-selected Milky Way halo
populations with highly polluted tracts of the M31 minor axis
\citep{Ibata07, Gilbert07}. In this study, we have carried out a
homogeneous analysis of 14 deep HST/ACS pointings, including the three
\cite{Brown06a} fields, which probe various regions in the inner halo
(R$_{proj}\lesssim 45$~kpc) and extended disk of M31. Many of these
fields were specifically targeted as bright and/ or unusual
substructure identified in the course of the INT/WFC imaging survey,
while others were taken from studies designed for other purposes.  In
this paper, we present a comparative analysis of these fields and
identify striking trends which indicate the inner halo is dominated by
a complex mix of giant stream material and disrupted disk material. A
future paper will present detailed SFH fits to these fields. In
Section~\ref{sec:obs} we describe our observations. The photometric
reduction and completeness testing are summarized in
Sections~\ref{sec:photom} and ~\ref{sec:compl}. The analysis and
results are presented in Section~\ref{sec:results}. Our interpretation
of the results is discussed in section~\ref{sec:discussion} and
summarized in section~\ref{sec:conc}.


\section{Observations}
\label{sec:obs}

The primary dataset analyzed here comes from our programs to obtain
deep imagery of prominent substructure in the M31 halo (GO 9458 and GO
10128, PI Ferguson).  Ten different regions in the M31 outskirts were
targeted in the course of these studies and nine of these are
presented in this paper (see Figure 1).  Preliminary results on the
stellar populations at six locations (`Giant Stream' (GS), `North-East
Shelf', `Minor Axis', `NGC 205 Loop', `Northern Spur' and `G1 Clump')
were previously presented in \cite{F05}. Each field was observed in
the F606W (broad V) and F814W (broad I) filters with the Wide-Field
Channel (WFC) of the ACS.  Most of the fields were observed for one
orbit in F606W and two orbits in F814W, resulting in accurate
photometry to several magnitudes below the horizontal branch. The
`Warp' field was exposed for significantly longer however the analysis
presented here is based on re-drizzled images to match the depth of
the other fields. The Minor Axis, NE Shelf, GS and NE Clump fields all
consist of two separate pointings to increase star count statistics on
otherwise low density stellar fields.

We also utilize two fields (`EC1\_field' and `GC6\_field') which were
observed as part of our imaging survey of newly discovered globular
clusters in the outer M31 halo (GO 10394, PI Tanvir, see
\cite{Mackay06, Mackay07}). These fields are somewhat shallower than
the primary fields discussed above (one orbit in each of F606W and
F814W) but they still provide good sensitivity to well below the
horizontal branch. With the globular clusters masked out (R $\sim$
30\arcsec), the background field populations are very useful probes of
diffuse structure in M31's extended disk. This is the first time
results from the Warp, NE Clump, Claw, EC1\_field and GC6\_field have
been published.

\begin{deluxetable}{c c c c c }[hb!]
  \tablewidth{8.0cm} \tablecaption{Distance and Reddening
    \label{tab:dist}} 
\tablehead{
    \colhead{Field} &
    \colhead{$E(B-V)$\tablenotemark{a}} & 
    \colhead{$N(HI)$\tablenotemark{b}}  &
    \colhead{$R_{proj}$\tablenotemark{c}} &
    \colhead{$R_{disk}$\tablenotemark{d}} \\
    &  &  \colhead{($10^{19} ~{\rm cm}^{-2}$)} &  \colhead{(kpc)} & \colhead{(kpc)}}

    \startdata
    Brown-spheroid & 0.081 &  0.65    & 11.5 & 53.0 \\
    EC1\_field     & 0.070 &  $<$0.10 & 13.2 & 60.5 \\
    GC6\_field     & 0.074 &  11.0    & 13.8 & 26.1 \\       
    NGC205 Loop    & 0.076 &  \nodata & 17.0 & 71.5 \\
    Minor Axis     & 0.060 &  0.64    & 17.9 & 91.8 \\
                   & 0.059 &  $<$0.10 & 19.9 & 82.7 \\
    NE Shelf       & 0.065 &  $<$0.10 & 18.6 & 54.3 \\
                   & 0.071 &  $<$0.10 & 19.3 & 59.5 \\
    G Stream       & 0.058 &  $<$0.10 & 19.0 & 79.5 \\ 
                   & 0.051 &  $<$0.10 & 20.7 & 68.4 \\ 
    Brown-stream   & 0.053 &  $<$0.10 & 20.3 & 72.7 \\ 
    Claw           & 0.060 &  1.20    & 23.7 & 42.0 \\
    Warp           & 0.054 &  9.80    & 25.1 & 31.1 \\
    N Spur         & 0.079 &  0.10\tablenotemark{e} & 25.3 & 43.3 \\
    Brown-odisk    & 0.080 &  35.0\tablenotemark{e}    & 25.6 & 25.6 \\
    G1 Clump       & 0.063 &  26.0                     & 29.2 & 29.7 \\
    NE Clump       & 0.093 &  1.20\tablenotemark{e}    & 44.1 & 59.0 \\
                   & 0.092 &  1.00\tablenotemark{e}    & 44.6 & 53.0\\
    \enddata 
    \tablenotetext{a}{Values from the reddening map of \cite{Schlegel98}.}  
    \tablenotetext{b}{M31 column densities; D. Thilker, private communication.}
    \tablenotetext{c}{Projected radial distance.}  
    \tablenotetext{d}{De-projected radial distance calculated assuming an inclined disk with PA$=38.1^{\circ}$ and $i=77.5^{\circ}$.} 
    \tablenotetext{e}{Line-of-sight subject to MW confusion.}
\end{deluxetable}

The remaining three fields in our study come from several of the
ultra-deep programs of Brown (GO 9453 and GC 10265). These pointings
are labelled `NGC224-Stream', `NGC224-Halo', and `NGC224-Disk' in the
HST archive, though for clarity here we reassign the names
`Brown-stream', `Brown-spheroid', and `Brown-odisk' respectively. For
each field, we retrieved a set of exposures from the HST archive and
drizzled them to an equivalent depth of our primary fields.  The
full-depth pointings have already been thoroughly explored and are the
subject of several papers \citep{Brown03, Brown06a, Brown06b}.  We
note that although the Brown-stream field is sandwiched between our
own GS pointings, we have chosen to keep it separate rather than
co-add it for comparative purposes.  The complete observational
information for each pointing analyzed here is given in
Table~\ref{tab:obs}.

\section{Data Reduction and Photometry}
\label{sec:photom}

All of the fields were subject to the same method of reduction and
analysis.  Pipeline-calibrated images were retrieved from the HST
archive and, for each field, those images in a given filter were
combined with the PyRAF TWEAKSHIFTS and MULTIDRIZZLE tasks
\citep{Koekemoer06}.  TWEAKSHIFTS was first used to calculate any
residual offsets between images after applying the WCS information
from the input headers. Such shifts were generally found to be
$\lesssim 0.5$ pixel in the case of images taken within a given visit
and $\lesssim 1$ pixel for images taken across different
visits. MULTIDRIZZLE was then invoked to register and stack images
using these corrected shifts. The final drizzles were conducted using
the Lanczos3 kernel with pixfrac and scale set to unity.

Before photometering the images, masks were constructed on the F814W
frames to cover diffraction spikes, bright background galaxies and
saturated stars and then applied to the F606W images. Photometry was
performed on the co-added images using the stand-alone version of
DAOPHOT-II \citep{Stetson87}.  The stellar density in our fields was
typically low to moderate. Tests revealed that tighter CMD sequences
generally resulted from aperture photometry alone however full
PSF-fitting photometry was also carried out as a means to discriminate
between stars and galaxies (see also \cite{F05,Faria07}). After a
first pass of object detection and aperture photometry, $\sim$250
candidate PSF stars were selected from the star catalogue on the basis
of their isolation, brightness and lack of obvious nearby
artifacts. In creating an empirical PSF for each field in each pass
band, any star whose fit was flagged as lying 1$\sigma$ or more from
the mean $\chi^2$ value was removed from the PSF candidate star list
and the PSF re-defined. The initial guess (non-varying) PSF model was
fit to all sources in the catalogue with ALLSTAR II.

Subsequently, all stars, except PSF candidate stars, were fit and
subtracted from the image and an improved PSF model was built and then
re-fit to the entire star catalogue. This process was iterated, allowing the
PSF model to vary linearly with position across the frame, until the
$\chi^2$ statistic of PSF candidate star fits was minimized and the
number of stars retained by ALLSTAR II had converged. A further pass
of object detection and PSF fitting photometry was carried out on a
star-subtracted frame to reveal stars previously missed. Finally, the
best-fit PSF model was fit to all stars in the catalogue. The
photometry list was pruned by rejecting stars with any of $\chi^2$,
magnitude error or sharpness lying outside 3$\sigma$ of the average
value at that magnitude, yielding a final list of stellar sources for
which we subsequently adopted their aperture photometry.  The two
catalogues (F814W and F606W) for each pointing were then matched to
within two pixels.  In total, approximately 675,000 stars were
retained for analysis.

We then corrected the photometry for aperture effects. The 2 to 4
pixel ($0.^{\prime\prime}05$ to $0.^{\prime\prime}10$) radii aperture
corrections were calculated using the average value from the
photometry of the pruned, matched list of PSF candidate stars. The 4
to `infinite' pixel aperture correction was taken from
\cite{Sirianni05}. Magnitudes were placed on the VEGAmag system
utilizing the zero-point values of \cite{Sirianni05} and corrected for
foreground reddening by interpolating within the maps of
\cite{Schlegel98} (see Table~\ref{tab:dist}). The photometry has not
been corrected for reddening internal to M31 however this correction
is not expected to be significant for most of our fields since they
lie outside the main gas disk.  Table~\ref{tab:dist} lists the
reddening, internal HI column densities, projected and de-projected
radial distances of each field and is ordered according to increasing
projected radial distance from the center of M31.  We did not correct
for charge transfer efficiency (CTE) because these data were obtained
early in the lifetime of the ACS when the cumulative damage caused by
radiation was minimal (see also \cite{Brown06a}). Hereafter, all
magnitudes refer to de-reddened magnitudes in the VEGAmag system.

\section{Completeness}
\label{sec:compl}

The morphology of a CMD reflects the fraction of stars in different
evolutionary phases; the precise location of a star within a given
phase is a function of both age and metallicity.  It is necessary to
have a good understanding of how complete the CMD is as a function of
magnitude and color before detailed interpretation. An artificial star
test algorithm was employed to empirically calculate the photometric
scatter and completeness of the data. Each image had several thousand
artificial stars of a specific magnitude (spanning the range 30.0 to
20.0 mag in 0.5 mag steps) added randomly, where each artificial star
is a scaled version of the best-fit PSF model. The aim is to add as
many artificial stars as possible to gain sufficient statistics, but
not so many as to significantly crowd the field. It was found that
adding $\sim10\%$ of the total number of real stars found per band
satisfied this condition in all fields.  The artificial stars are
recovered, with no prior knowledge regarding their positions, using
the identical method employed for the real data. They are pruned using
the same parameters derived from pruning the original photometry, not
least since the excess artificial stars at the magnitude being tested
could skew the 3$\sigma$ levels used in the pruning process. The
resulting catalogue is compared with the actual input positions of the
artificial stars in order to derive the success rate of
recovery. Steps are taken to avoid the accidental inclusion of real
stars close to, or superimposed on, artificial star
positions. Artificial stars coincident with a brighter real star are
rejected, while artificial stars coincident with fainter real stars
are retained but considered to have contaminated photometry. The
process is repeated until the entire magnitude range has been covered.

\begin{figure}
\begin{center}
\includegraphics[width=6.5cm, angle=90]{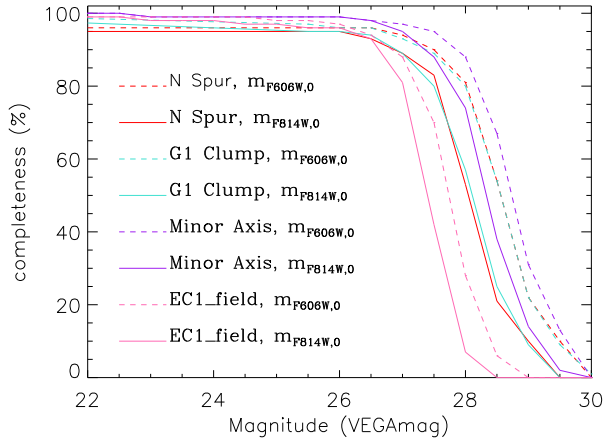}
\end{center}  
\caption{Results of the artificial star tests on representative fields
of high, low and intermediate stellar crowding as demonstrated by the
variation in completeness as a function of magnitude for the Spur,
Minor Axis and G1 Clump respectively. Also shown is the completeness
function for the EC1 field which, along with GC6, has a reduced
exposure time relative to the other fields analyzed here.}
  \label{fig:compl}
\end{figure}

Figure~\ref{fig:compl} shows the completeness rate as a function of
magnitude for fields spanning a range of stellar densities, as well as
for one of the shallower fields.  For the bulk of our fields,
completeness levels exceed 80$\%$ at $m_{F606W}=28.0$ mag and
$m_{F814W}=27.5$ mag. For the slightly shallower EC1\_field and
GC6\_field, these limits drop to 27.25 mag and 27.0 mag
respectively. Thus, a high completeness is maintained to generally
$\gtrsim$ 2 magnitudes below the RC.  As expected, the higher density
fields have somewhat lower completeness at a given magnitude. Analysis
of the photometric errors as a function of magnitude and color as
derived from artificial star tests demonstrate that magnitudes are
accurate to $\lesssim$0.10 magnitudes at $m_{F814W}$ = 26.5 mag.


\section{Analysis and Results}
\label{sec:results}

\subsection{Color-Magnitude Diagrams}

\begin{figure*}
\begin{center}
\includegraphics[width=16.0cm]{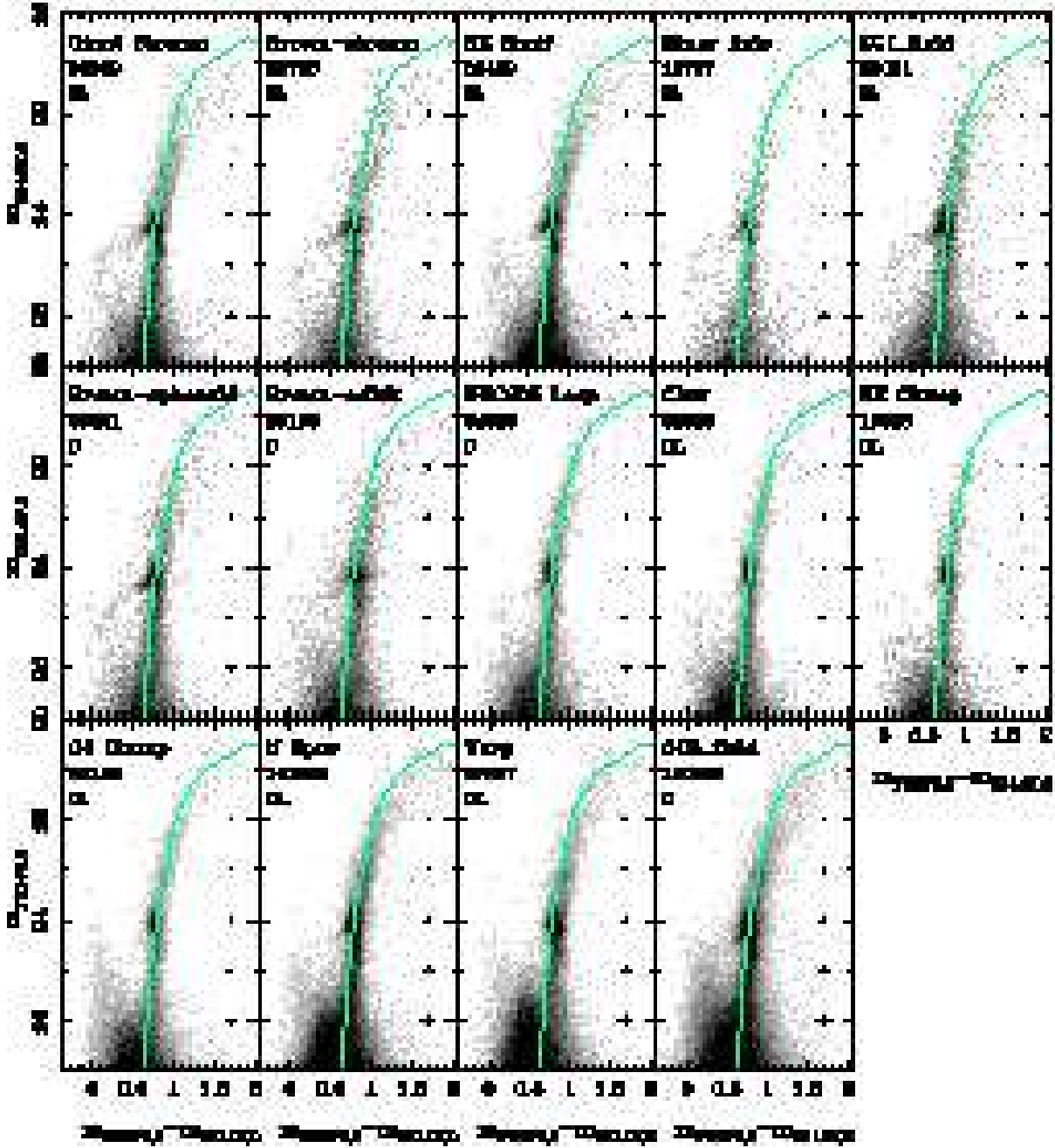}
\end{center}  
\caption{Color magnitude diagrams plotted as Hess diagrams with a
  square-root stretch to bring out fainter features such as RGB and
  AGB bumps. The bins are 0.025 mag in color and 0.062 mag in
  magnitude. The number of stars in each diagram is given below the
  name. The ridge line of 47 Tuc, which has [Fe/H] = -0.7 and age =
  12.5 Gyr \citep{Brown05}, has been shifted to the distance of M31
  and over-plotted. Photometric errors derived from artificial star
  tests are shown on the right-hand side of each CMD. The CMDs are
  $\gtrsim$ 90 $\%$ complete at $m_{F814W,0} = 27.0$mag ($\sim$ 80
  $\%$ for EC1\_field and GC6\_field). Fields are labelled as
  disk-like (DL), stream-like (SL) or composite (C) as described in
  the text (Sec 5.2).}
  \label{fig:hess}
\end{figure*}

Figure~\ref{fig:hess} shows the CMDs of all of the fields represented
as Hess diagrams with a square-root stretch. The total number of stars
detected in each field is given below the name, and error bars are
plotted as derived from artificial star tests. The ridge-line of 47
Tuc (NGC104: [Fe/H] = -0.7, 12.5 Gyr) as taken from \cite{Brown05} is
superimposed on the CMDs. It has been transferred into the VEGAmag
photometric system using the appropriate zero-points from
\cite{Sirianni05}, corrected to the M31 distance modulus of (m-M) =
24.47 \citep{McC05} and de-reddened. The CMDs appear broadly similar;
all have wide RGBs and prominent RCs typical of M31 populations seen
at smaller radii \citep[e.g.,][]{Mould86, Holland96, FJ01,
Bellazzini03, F05, Brown06a}.  The RGB widths are considerably larger
than the photometric error bars for magnitudes brighter than
$m_{F814W}<$ 26 mag suggesting an intrinsic spread in metallicity. The
high proportion of RGB stars with colors redder than the 47 Tuc
ridge-line suggests that these stellar populations extend to high
metallicities ([Fe/H] $\gtrsim$ -0.7 dex). The Warp, GC6\_field and
Brown-odisk fields reach further to the red than other fields. Since
these fields are amongst those with the highest HI column densities
(see Table~\ref{tab:dist}), it is possible that internal extinction is
also contributing to the color of the RGB.

On closer inspection, several morphological differences are apparent
between the CMDs in Figure~\ref{fig:hess}. The most striking of these
is the morphology of the RC and horizontal branch (HB) features
produced by core Helium burning stars. In many of the CMDs, the RC
slants blueward at fainter magnitudes (hereafter referred to as the
blue RC\footnote[9]{Often referred to as a red horizontal branch in
the literature.}) and is accompanied by an extended horizontal
branch. In contrast, many of the other CMDs show fairly round RCs
which have a significant luminosity but little color spread and no
evidence for an extended horizontal branch. Additionally, these fields
also show evidence for blue plumes (BPs) of upper main-sequence (i.e
young) stars. If extended blue HBs are also present in these fields,
they would partly overlie the BPs making direct detection difficult.
One further characteristic common to many of the latter set of fields
is a prominent overdensity to the left of the RGB at $m_{F814W} \sim$
26.1 mag.  This feature has the correct magnitude and color to be be a
main sequence turn-off of a 2-3 Gyr population \citep{Girardi00}.  For
reasons which will become apparent later on, we assign the terms
`stream-like' and `disk-like' respectively to these different CMD
behaviors.  Note that while CMDs within a given group share a strong
morphological resemblance, they need not be composed of exactly
identical populations. Indeed, the average best-fit age and
metallicity of the Brown-stream and Brown-spheroid fields, both
classified here as `stream-like', have been found to differ by 0.1 dex
and 0.9 Gyr respectively \citep{Brown06a}.

Well populated fields in both categories exhibit a prominent RGB bump
located below the RC ($m_{F814W} \sim$ 24.6 mag, $m_{F606W}-m_{F814W}
\sim$ 0.8 mag) implying a metal rich population. These features are
the result of stars zig-zagging past the same luminosity value when
the H burning shell passes the chemical discontinuity left behind by
the expanding convective envelope causing a temporary drop in nuclear
efficiency. The RGB bump luminosity is expected to decrease with
respect to the RC as the metallicity increases though it can be
difficult to interpret in the case of composite stellar
populations. The most populated fields (NE Shelf, N Spur GC6\_field
and Warp) also feature asymptotic giant branch (AGB) bumps at
$m_{F814W} \sim$ 23.1 mag, $m_{F606W}-m_{F814W}\sim$ 0.9 mag. For
fields with both RGB and AGB bumps, the fact that their luminosities
straddle the RC luminosity supplies an additional clue regarding their
ages and metallicities. According to the RC models of \cite{Alves99},
such behavior indicates that these populations are relatively metal
rich. Further, for a single stellar population of [Fe/H] = -0.7 dex,
this behavior implies an age of at least $\sim$5 Gyr, however this
restriction is relaxed for more metal rich populations

\begin{figure*}[ht!]
\begin{center}
\includegraphics[width=16.0cm]{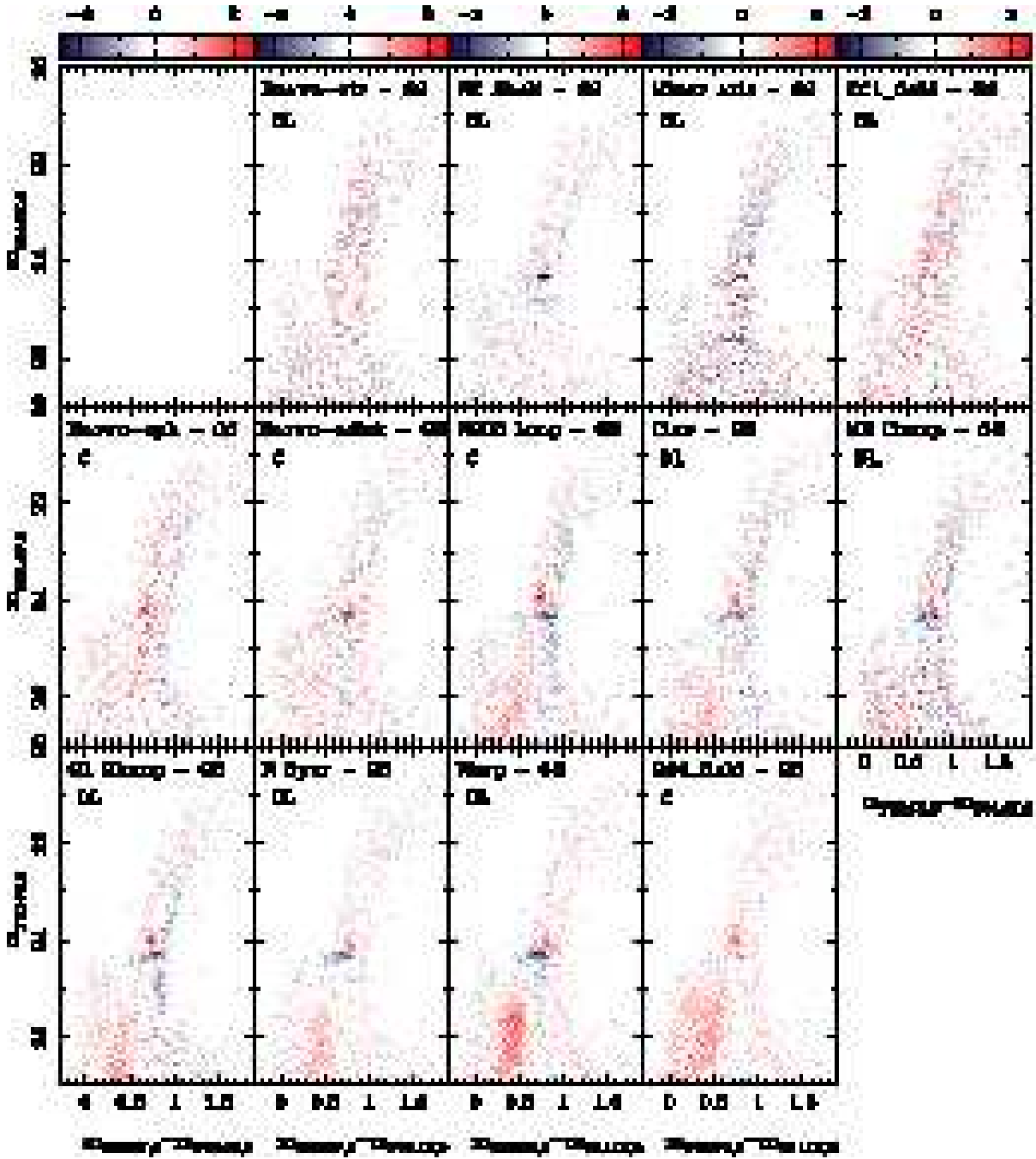}
\end{center}
  \caption{Comparisons of individual fields to the GS field. In
each panel, the normalized Hess diagram of the field has been
subtracted by the normalized GS Hess diagram to highlight the relative
differences in the underlying populations. Red signifies an
under-subtraction and blue an over-subtraction. The units are
arbitrary and meaningful differences are strong, coherent
overdensities in red or blue. Subtractions involving the Minor Axis,
Brown-stream, EC1\_field and NE Clump fields are noisy due to small
number statistics.}
\label{fig:stream}
\end{figure*}

\begin{figure*}[ht!]
\begin{center}
\includegraphics[width=12.5cm,angle=90]{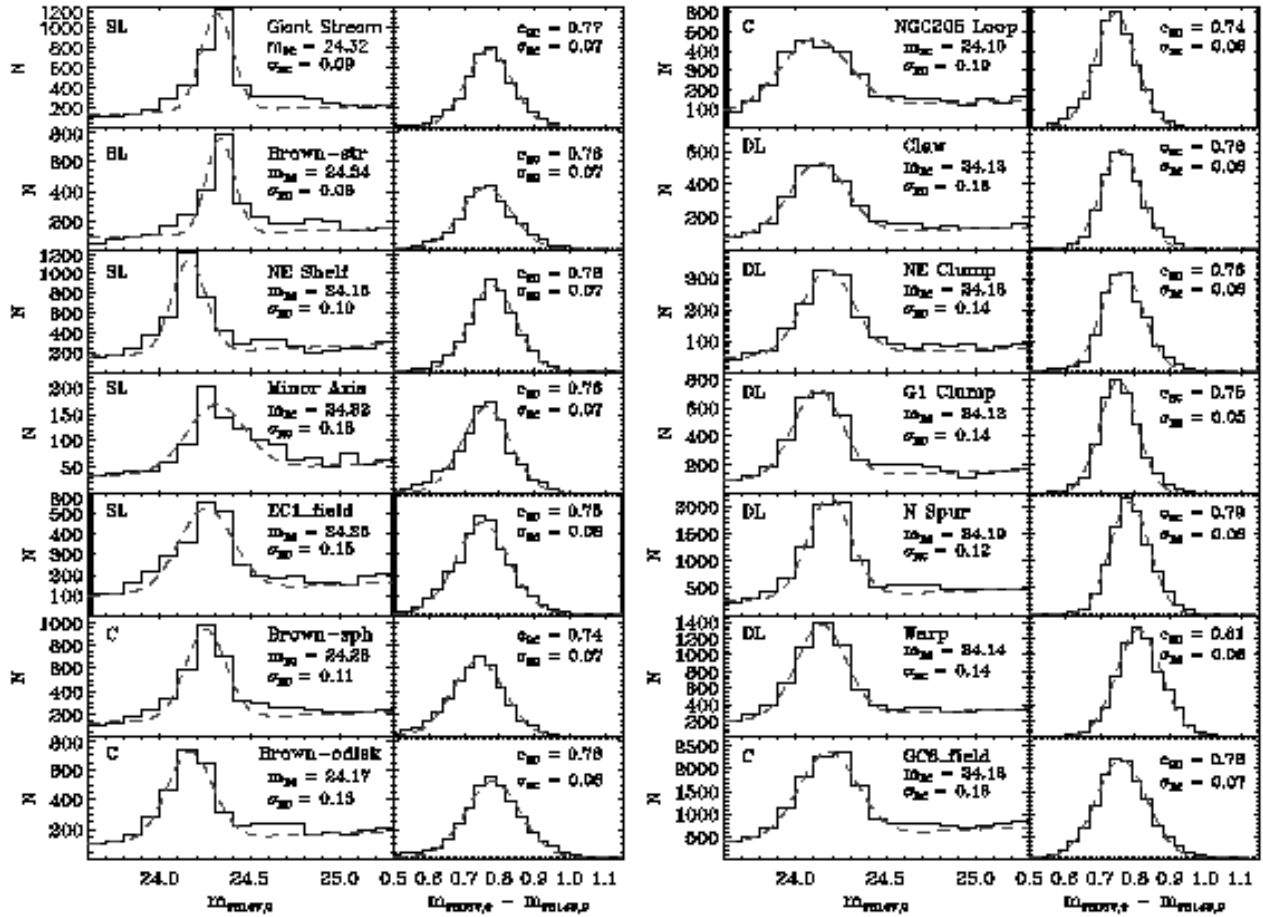}
\end{center}
  \caption{Gaussian fits (dashed line) to the red clump $m_{F814W}$
  luminosity function (\textit{left panels}) and color distribution
  (\textit{right panels}). $m_{RC}$ refers to the peak $m_{F814W}$
  magnitude and $c_{RC}$ to the peak $m_{F606W}-m_{F814W}$
  color. Errors in the fits are $\pm$ 0.01 in all cases except for the
  Minor Axis which has an error of $\pm$0.02 in both $m_{RC}$ and
  $c_{RC}$.}
\label{fig:gauss}
\end{figure*}

A few fields exhibit CMDs which share characteristics of both
stream-like and disk-like behavior. For example, the Brown-odisk field
has a round RC typical of disk-like fields, yet it also shows evidence
for a low-level blue RC and even an extended HB.  Another field with
some overlap between the two groups is the NGC205 Loop. As noted in
\cite{F05},this field has a blue RC but it lacks the strong RGB and
AGB bumps seen in other stream-like fields. This is not just a
statistical effect related to the stellar density on the CMD because
the RGB bump is noticeably stronger in the Brown-stream field which
has $\sim$11,000 fewer stars. In addition, there is a hint of the
bright main-sequence turn-off `hump' ubiquitous to our disk-like
fields. In Figure 3, the Brown-spheroid appears very similar to the GS
despite being presented as pure halo in the literature
\citep{Brown06a,Brown06b}. However, it will be shown in the following
sections that there are important differences between these fields and
that its true nature is likely a mixture of both GS and halo
populations. These fields thus form a third category of `composite
fields'.

\subsection{Differencing the CMDs}

Figure~\ref{fig:stream} shows the result of subtracting the normalized
GS Hess diagram from the other fields, where the Hess diagrams have
been normalized by the total number of stars with F814W band
completeness over 90$\%$ ($m_{F814W}< 27.0$ mag; corresponding to
$\sim$80$\%$ in the shallower GC6\_field and EC1\_field). As expected,
the stream-like fields show excellent agreement with the GS leaving
negligible residual structure, except in the case of the NE Shelf
where the RC is brighter. \cite{F05} have argued that this offset in
RC magnitude is the effect of different line-of-sight distances; we
will return to this point in Section~\ref{sec:streamdisc}.  Using
their ultra-deep CMD of the giant stream, \cite{Brown06a} have derived
an average age of 8.8 Gyr and average metallicity of [Fe/H] = -0.7 dex
in their field. It is therefore likely that this type of population
dominates all the stream-like fields.

The subtractions reveal very clear differences between the disk-like
fields compared to the GS and simple line-of-sight distance shifts
cannot bring the two into agreement. Disk-like fields display a
significant over-subtraction of the blue RC (at $m_{F814W} \sim$ 24.3
mag, 0.4 $< m_{F606W}-m_{F814W} <$ 0.9) and an equally prominent
under-subtraction of the vertically elongated RC. In addition, they
reveal varying degrees of under-subtraction of the blue main sequence
populations ($m_{F606W} - m_{F814W} \leq$ 0.6 mag, $m_{F814W} \geq$
25.5 mag) indicating that these fields have much greater proportions
of stars with younger ages than the GS. This residual young population
is especially strong in the Warp field. Several disk-like fields (the
Claw, NE Clump, G1 Clump and N Spur) also show a noticeable
over-subtraction redwards of their RGBs seen as a thin blue strip
running parallel to the RGB for $m_{F814W} < 25.5$ mag suggesting that
GS RGB stars extend to higher metallicities than these disk-like
fields (note that this over-subtracted band is too wide to be caused
by age differences alone).

The composite fields show differing patterns of residuals, none of
which exactly mimic those seen in the disk-like fields. Despite the
NGC205 Loop CMD showing a blue RC, there is a strong over-subtraction
of that feature in Figure~\ref{fig:stream} indicating that old
metal-poor stars constitute a much smaller percentage of this
population than in the GS. Indeed the subtraction has revealed that
the NGC205 Loop field has some disk-like properties; a RC elongated in
luminosity, a `hump' consistent with bright main-sequence turn-off
stars ($m_{F814W} \gtrsim $ 25.5 mag, $m_{F606W} - m_{F814W} \lesssim$
0.6 mag) and an over-subtraction redwards of the RGB. In terms of the
subtracted Hess diagram, the NGC205 Loop shows somewhat similar
residuals to the disk-like Claw field, although the under-subtracted
RC is slightly bluer.  The GS subtracts rather cleanly from the
GC6\_field although residuals confirm that it has an excess of younger
and more metal-rich constituents. The Brown-odisk field exhibits
smaller residuals in the region of the RC/HB than disk-like fields,
implying the possibility of a stream-like contribution to the
population in this region.

The last composite field, the Brown-spheroid, displays the most unique
subtraction signature. Although parts of the CMD are well matched by
the GS, the red side of the RGB is strongly over-subtracted suggesting
that this field does not contain such high metallicities. Furthermore,
the pattern of RC residuals reveals an excess HB sequence not seen in
the GS, or any other fields.  In figures throughout the paper, we will
refer to fields as being either DL (disk-like), SL (stream-like) or C
(composite) based on the Hess diagram subtractions.

\subsection{The Intermediate / Old Populations}
\label{sec:old}

The Hess diagram subtractions have confirmed that the dominant
difference between stream-like and disk-like CMDs are those features
which reflect the composition of the old and intermediate-age core
Helium burning components. We proceed here to put these differences on
a more quantitative ground.  Figure~\ref{fig:gauss} shows the
$m_{F814W}$ luminosity function (LF) and color distribution in the
region of the RC for each field. The region selected for the color
distribution plots was $23.5 < m_{F814W} < 24.75$ and $-0.5 <
m_{F606W} - m_{F814W} < 1.2$ to avoid the inclusion of BP stars. The
dashed line represents the Gaussian fits to these distributions, using
the fitting function formalism of \cite{PaczynskiStanek98}. At
$m_{F814W}\sim$24.0 mag, the typical peak RC magnitude, the data are
well above 90$\%$ complete and the photometric errors in magnitude
($\lesssim$ 0.02 mag) and color ($\lesssim$0.03 mag) are small.

The RC absolute magnitude and color are a function of the age and
metallicity of a population. For a given metallicity, old stars form a
fainter RC than young stars. The presence of a bright, elongated RC
signifies a predominantly metal-rich intermediate/young aged
population \citep{G&S01}.  For a population of known age and
metallicity, the RC magnitude also acts as a standard candle, a fact
we will exploit in Section~\ref{sec:streamdisc}.
Figure~\ref{fig:gauss} further indicates that the disk-like fields
generally have brighter and slightly more elongated RCs than the
stream-like fields, suggesting that they contain younger stellar
populations in addition to the intermediate/old RGB components. In the
absence of distance effects, the brighter peak magnitude could be the
result of a contribution from the more evolved counterparts of the
blue plume. That is, stars with the same age and metallicity but
slightly larger masses than the young main sequence turn-off stars
which causes them to evolve into RGB stars more rapidly.  The color
distributions of the RCs are very similar, regardless of grouping,
though there is marginal evidence for stream-like fields having
slightly broader color widths. The peak magnitude and color of the G1
Clump ($m_{F814W} = 24.12$ mag, $m_{F606W}-m_{F814W} = 0.75$ mag)
measured here are in excellent agreement with \cite{Faria07}.

As a test of the robustness of our RC measurements, as well as our
photometry in general, we measured the RC color and magnitude on each
individual CMD in those fields where we obtained double pointings (the
GS, NE Shelf, NE Clump and Minor Axis). The peak RC magnitudes and
colors for the two separate pointings, as well as for our GS field and
Brown-stream, are generally in excellent agreement, differing by no
more than $\pm0.03$ mag.  This indicates that the RC can be used as a
distance indicator to an accuracy of $\pm 11$~kpc.  An unusual result
was recovered in the case of the Minor Axis. In that field, the peak
RC magnitudes differed by a sizeable 0.10 mag in the two different
pointings. If the populations in these two pointings are identical,
this magnitude variation would signify a puzzling $\sim$ 38.5 kpc
line-of-sight difference despite a mere $\sim 2$ kpc projected
distance between them. More likely, we suspect we have serendipitously
discovered some small-scale substructure in this region with
intrinsically different populations. The very low stellar density in
the individual fields (combined these fields have $<$ 17,000 stars)
precludes a detailed assessment of the similarity of the two CMDs,
however no obvious differences are apparent. It is interesting,
however, that the inner of the two pointings contains twice as many
stars as the outer.

\begin{figure}
\begin{center}
\includegraphics[width=6cm, angle=90]{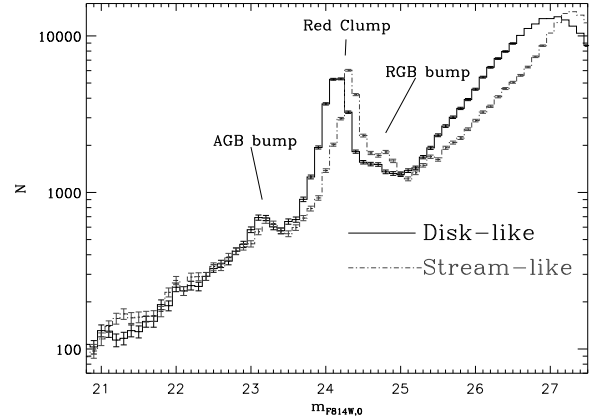}
\end{center}
  \caption{Comparing the completeness-corrected stream-like and
  disk-like RGB luminosity functions. The luminosity functions have
  been normalized at $m_{F814W}$ = 23.4 mag, between the RC and AGB
  bump.}
\label{fig:lumfunc}
\end{figure}

\begin{figure*}[ht!]
\begin{center}
\includegraphics[width=10.5cm, angle=-90]{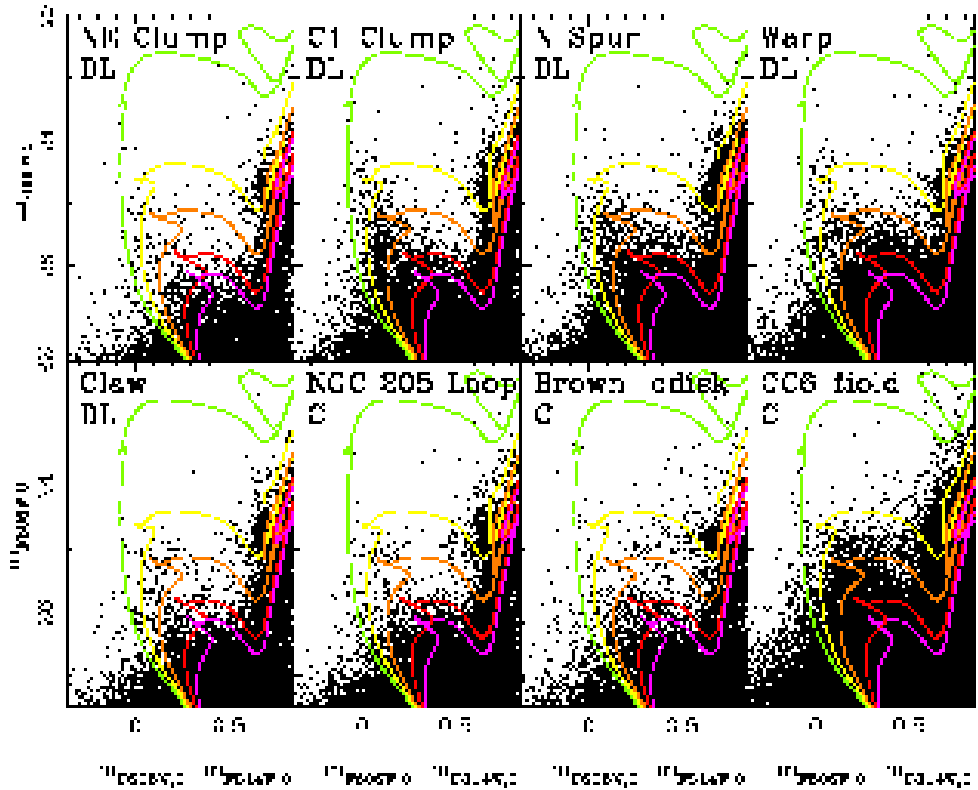}
\end{center}
  \caption{Close up view of those fields containing blue plume
populations.  Isochrones from the \cite{Girardi00} library of
metallicity [Fe/H] = -0.4 dex and ages 230 Myr (green), 630 Myr
(yellow), 1.0 Gyr (orange), 1.6 Gyr (red) and 2.0 Gyr (magenta) are
over-plotted. The density of stars falling between adjacent isochrones
gives an estimate of the relative strength of star formation in that
age bracket.}
\label{fig:bp}
\end{figure*}

Figure~\ref{fig:lumfunc} compares the completeness-corrected RGB
luminosity functions of stream-like and disk-like fields over the full
luminosity range.  The `stream-like' curve includes the GS,
Brown-stream, NE Shelf, Minor Axis and EC1\_field while the
`disk-like' curve includes the Claw, NE Clump, G1 Clump, N Spur and
Warp. Composite fields are omitted. Individual fields in each group
have first been shifted so that their peak RC magnitude matches the
average of that group before being co-added and the LFs have been
normalized at $m_{F814W}$ = 23.4 mag, a point mid-way between the RC
and AGB bump. This plot further emphasizes the gross differences
between the two populations, including the different RC morphologies,
with the disk-like function peaking at brighter magnitudes. In
addition, stream-like populations appear to have a higher proportion
of RGB bump stars than disk-like populations although the AGB bumps
are similarly concentrated. For magnitudes fainter than the RC, the
disk-like LF rises more steeply than stream-like LF highlighting their
young components as bright main sequence turn-off stars merge into the
RGB.

\vspace{0.5cm}
\subsection{The Young Populations}

Figure~\ref{fig:hess} indicates that the disk-like fields, and several
of the composite fields have sub-populations of young stars.  Full
CMD-fitting is required to place rigorous constraints on the recent
star formation history of these fields but we gain some quantitative
insight by comparing their BPs with theoretical isochrones in
Figure~\ref{fig:bp}. The \cite{Girardi00} isochrones of [Fe/H] = -0.4
dex with ages of 230 Myr, 630 Myr, 1.0 Gyr, 1.6 Gyr and 2.0 Gyr best
match the BP magnitude and color distribution. Lower metallicity
isochrones are too blue to match the BP while more metal rich
isochrones are generally too red except in the cases of the GC6\_field
and Warp where there is some overlap.  Several fields show BPs which
reach $m_{F606W}\sim$ 23 - 24 mag, indicative of stars as young as a
couple of hundred megayears old. It is particularly remarkable that
such a young population is seen in the NE Clump, a field which sits at
$\sim 44 ~{\rm kpc}$ (or $R_{disk}\sim 56 ~{\rm kpc}$ ).

In field populations, blue stragglers are thought to form by mass
transfer between primordial binaries \citep{Davies04}. These old blue
straggler stars can masquerade as young main sequence stars on a CMD
\citep{Carney05} so it is important to address whether the BPs we see
could be caused by such stars instead. \cite{Faria07} provide a
thorough discussion of this issue for the specific case of the G1
Clump and find that while this scenario cannot be completely ruled
out, it is highly unlikely; these same arguments apply to all the
fields presented here. If the BPs in our CMDs were made entirely of
blue stragglers, the naive expectation would be that denser fields
would have correspondingly dense BPs which is not observed. We also
note that disk-like fields consistently have higher HI column
densities than stream-like fields, further supporting the presence of
genuinely young stars in these regions.

\vspace{0.4cm}

\section{Discussion}
\label{sec:discussion}

\subsection{The Nature of the Stream-Like Fields}
\label{sec:streamdisc}

The discovery of the giant stream \citep{Ibata01a} was the first
spectacular indication that M31 is still accreting mass. Fields in our
study which directly probe this substructure are the GS and
Brown-stream fields, both lying at $\sim 20 ~{\rm kpc}$ projected
radial distance from M31 (see Figure~\ref{fig:map}). In addition, we
have shown that the NE Shelf, EC1\_field and Minor Axis fields have
CMDs which are strikingly similar to the GS fields, indicating that
these fields are also dominated by tidal debris stripped from the GS
progenitor.  Although these fields generally have smaller projected
radii than disk-like fields, their positions correspond to larger
distances in the plane of the disk.

\begin{figure}
\includegraphics[width=8.5cm]{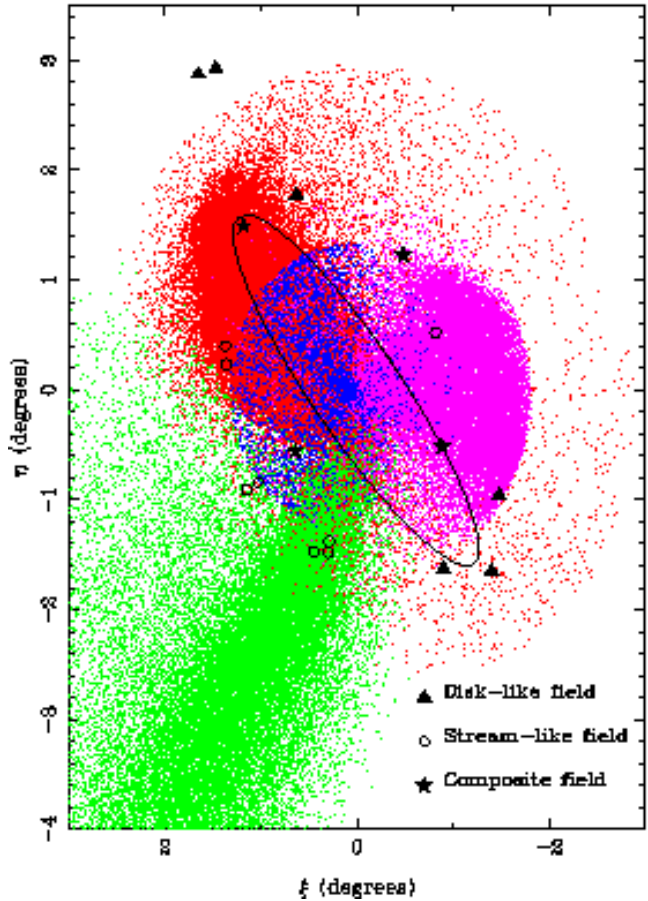}
  \caption{Locations of our fields with respect to the Fardal et
  al. (2007) simulation showing the predicted distribution of GS
  progenitor tidal debris around M31. Color coding is as follows:
  \textit{green} particles represent the GS falling into its first
  pericentric passage; \textit{red} particles show the fanning out of
  material into the Northeast shelf after the first close pass and
  approaching their second pericentric passage; \textit{magenta}
  particles show the fanning out of material into the Western shelf
  after the second close pass and approaching their third pericentric
  passage and \textit{blue} particles are heading for their forth
  pericentric pass. The locations of our 14 HST/ACS fields are
  over-plotted (see Figure 1 for labels). The ellipse shows the
  approximate extent of M31's bright disk ($R = 27$~kpc,
  PA$=38.1^{\circ}$ and $i=77.5^{\circ}$). The overlap between
  stream-like fields (open circles) and predicted coverage of stream
  debris is striking. Disk-like fields are represented by filled
  triangles and composite fields by filled stars.}
 \label{fig:fardal}
\end{figure}

\cite{Fardal07} (hereafter F07) have recently presented an N-body
simulation of the accretion of a dwarf satellite of mass $\sim 10^9$
~M$_{\sun}$ within a realistic M31 potential. Their simulation has
been tailored to reproduce the giant stream and North-east shelf as
observed in Figure 1. As the satellite falls towards the galaxy, long
tidal streams are produced both leading and trailing the progenitor
core. They identify the trailing stream as the giant stream and
suggest that the leading stream may have wrapped around the inner
galaxy at least twice.  Figure~\ref{fig:fardal} shows the predicted
distribution of the GS progenitor debris in their model with our
HST/ACS pointings overlaid. Note that only satellite particles are
represented; inside the black ellipse (which indicates the extent of
M31's bright disk) a significant contribution of M31 disk stars is
expected.  As can be seen, there is an excellent overall agreement
between the stream-like fields identified in this paper and the
predicted pattern of debris. In particular, this includes the
``Western Shelf'' overdensity which is probed for the first time by
our EC1\_field. Figure~\ref{fig:fardal} also indicates that much of
the minor axis of M31 should be contaminated by diffuse GS debris
stripped off during various pericentric passages (see
\citealt{Gilbert07}) and agrees with our finding that both the Minor
Axis and Brown-spheroid fields have significant stream-like
components.  It is also encouraging that our disk-like fields, with
the exception of the Claw field which projects on the very edge of
Fardal's Western Shelf, are found to lie well away from regions that
contain stream debris in the simulation.

The three-dimensional distribution of stream debris is of great
importance for placing constraints on the progenitor orbit
(e.g. \cite{McC03}).  \cite{F05} and \cite{Brown06b} have exploited
the similarity of the populations in stream-like fields to determine
differential distances based on RC magnitudes. In particular,
\cite{F05} applied this to the GS and NE Shelf fields and derived that
the latter lay closer to us than the former by a factor of
$1.07\pm0.01$. The observed RC magnitude is distance dependent, so if
we assume that the underlying populations in stream-like fields are
the same, we can examine their differential line-of-sight distances by
comparing their peak RC magnitudes.  Table~\ref{tab:los} presents a
summary of the distance measurements to stream-like fields identified
here.  The observed distances have been determined by assuming that
the GS field lies at ${\rm D_{los}} = 830 \pm 20 ~{\rm kpc}$
\citep{F05} with respect to M31 (${\rm D_{los}}$ = $785 \pm 25 ~{\rm
kpc}$; \cite{McC05}). The RC magnitude differences between the GS and
other fields then yield relative distances from the center of M31.
The line-of-sight depth of simulated particles is taken from Figure 2b
of F07, set on an absolute scale using the same ${\rm
D_{los}(M31)}$. There is an excellent agreement in both the
differential and absolute line-of-sight distances.

\begin{deluxetable}{c c c c c }[h!]
  \tablewidth{8cm} \tablecaption{Line-of-Sight Distances to Stream-like Fields
    \label{tab:los}} 
\tablehead{
    \colhead{Field} &
    \colhead{${\rm D_{los}^{obs  }}$\tablenotemark{a}} & 
    \colhead{${\rm D_{los}^{sim  }}$\tablenotemark{a}}  &
    \colhead{$F_{GS/field}^{obs}$\tablenotemark{b}} &
    \colhead{$F_{GS/field}^{sim}$\tablenotemark{b}} \\
    &  \colhead{$\pm 20~{\rm kpc}$} & \colhead{}  &  \colhead{$\pm 0.03$} & \colhead{}}

    \startdata
    Giant Stream   & 830 & 816  &   -  &   -  \\
    EC1\_field     & 807 & 795  & 1.03 & 1.03 \\
    Minor Axis     & 785 & 785  & 1.06 & 1.04 \\       
    NE Shelf       & 769 & 745  & 1.08 & 1.10 \\
    \enddata 

    \tablenotetext{a}{Observed and simulated ${\rm D_{los}}$
      calculated assuming M31 lies at ${\rm D_{los}} = 785 \pm 25
      ~{\rm kpc}$ \cite{McC05}).}
    \tablenotetext{b}{Factor, $F_{GS/field}$, by which the GS lies
      further away from us than the field in question.}
\end{deluxetable}

\begin{figure}[h]
\includegraphics[width=8cm]{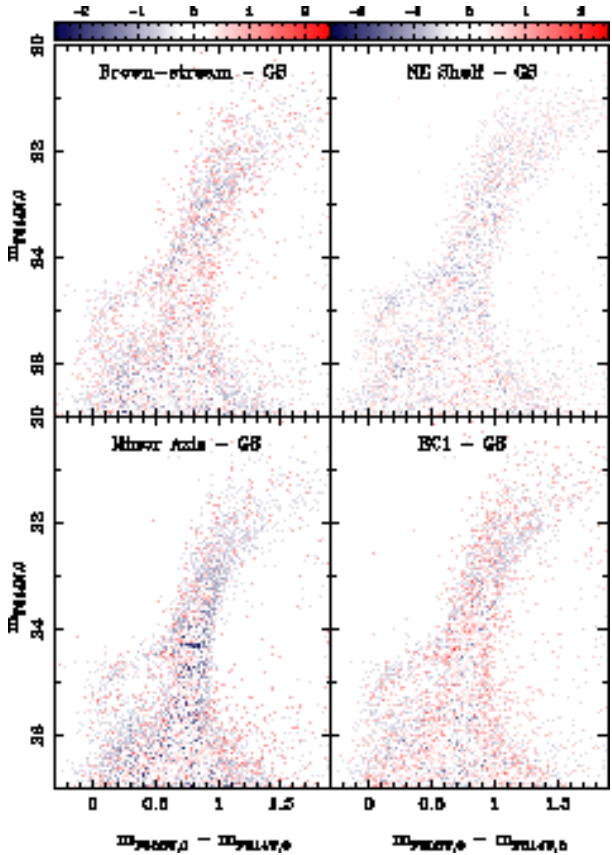}
  \caption{Same as Figure~\ref{fig:stream}, but with stream-like
  fields shifted to the same line-of-sight distance as the GS field
  using RC color and magnitude information.}
 \label{fig:shift}
\end{figure}

In using RC brightnesses to calculate approximate line-of-sight
differences, we have implicitly assumed that all the stream-like
fields have very similar underlying populations. To test the accuracy
of this assumption, we have used the mean magnitude and color of the
RC to shift all of the stream-like fields to the same line-of-sight
distance and reddening as the GS field and then re-subtracted the GS
Hess diagram (see Figure~\ref{fig:shift}). These subtractions are much
cleaner than those in Figure~\ref{fig:stream} and the residuals are
consistent with Poisson noise. This lends support to the idea that
they are composed of the same mix of stellar populations as the GS
field. The Minor Axis retains a slight over-subtraction in the RC
area, but it is unclear how significant this is given the sparsity of
this field. Moreover, as we saw in \S~\ref{sec:old}, there may be
small-scale population variations within this field.

\subsection{The Nature of the Disk-Like Fields}

Aside from sharing a common RC morphology, the disk-like fields are
also distinguished by the presence of a young ($\leq 2~{\rm Gyr}$)
population, the prominence of which varies from field to field. The NE
Clump, G1 Clump, Spur, GC6\_field and Warp appear to have formed stars
as recently as a few hundred Myr ago. There is no clear correlation
apparent between the strength of the BP population and either the
projected or deprojected radial distance of the field, perhaps
unsurprising given that the gas disk in M31 is known to be non-planar
at large radius \citep{BrinksBurton84}. However, there is a clear
correlation between the presence of BP stars and the local HI
column. For example, the fields with the highest unconfused HI column
densities are the Warp ($N_{HI}=9.8 \times 10^{19} ~{\rm cm^{-2}}$)
and the G1 Clump ($N_{HI}=26.0 \times 10^{19} ~{\rm cm^{-2}}$) which
also exhibit the strongest BPs. This could suggest that the young
stars seen have formed in situ.

It should be noted that disk-like fields are roughly aligned with the
major axis and many (G1 Clump, N Spur, Warp, Claw and the NE Clump, in
addition to the NGC205 Loop) probe well-defined high surface
brightness substructure lying within the extended disk of
\cite{Ibata05}. This intriguing structure is characterized by strong
rotation in the distance range $15 ~{\rm kpc} \leq R \leq 40 ~{\rm
kpc}$, slightly lagging the thin disk component by $\sim 40 ~{\rm
km~s^{-1}}$, and has a modest velocity dispersion of $30~{\rm
km~s^{-1}}$. \cite{Ibata05} have discussed various scenarios for the
formation of the extended disk and concluded the most likely one is
via accretion.  We revisit this important question here in light of
the new stellar populations constraints we have derived.

\subsubsection{Could disk-like material originate from one or more accreted satellite galaxies?}  

The idea that the disk-like material results from the recent accretion
of many small satellites provides a ready explanation for the jumbled
morphology of the substructure.  However, the homogeneity of the
populations in the various fields -- which are widely separated in
distance in some cases -- is problematic.  Within the Local Group low
mass satellite galaxies generally obey a mass-metallicity relation and
exhibit a range of star formation and chemical evolution histories
\citep[e.g.,][]{Dolphin05}.  Models which explicitly take this into
account predict that stellar halos should possess significant age and
metallicity inhomogeneities resulting from accretion events with a
range of satellite masses \citep{Font06, Font07}.  As we have
discussed, such differences are not observed in the disk-like
substructure. Moreover, \cite{Font07} have shown that young
populations are not expected in tidal debris lying within the inner
($< 50~{\rm kpc}$) regions of hierarchically built halos.

An alternative scenario is one in which the disk-like material comes
from a single accretion event involving a more massive satellite.
\cite{penarrubia06} have shown that an extended exponential disk
structure can result from the accretion of a $10^9-10^{10}$~M$_{\sun}$
object on a coplanar prograde object. However, in order to reproduce
the high degree of rotation observed in M31's extended disk, they
require the progenitor to start on a circular orbit which is rather
atypical for satellites observed in cosmological simulations
\citep{vandenBosch99}.  Although the homogeneity of the underlying
stellar population is not a problem in this scenario, the presence of
genuinely young stars is rather puzzling. Indeed, this would seem to
require that the disrupting satellite retain a substantial reservoir
of gas until well within the potential of M31, even while much of its
stellar component has already been stripped.

Could the source of disk-like substructure be the GS progenitor
itself? As shown in Table~\ref{tab:dist}, disk-like fields have HI
column densities at least an order of magnitude larger than
stream-like fields.  It could be that the young populations in these
fields have formed in situ and are unrelated to the dominant stellar
component which has come from the stream.  Our results discount this
possibility. We have shown that the old and intermediate aged stars in
stream-like fields have age and metallicity distributions which cannot
be reconciled with those in disk-like fields, as reflected by their
differing RC and HB morphologies.  Furthermore, stream-like fields
appear to have been more chemically enriched than disk-like fields at
least a few gigayears before the present time. Current models predict
`first contact' between the GS progenitor and M31 around $\sim 1 ~{\rm
Gyr}$ ago \citep[e.g.,][]{Ibata04, Fardal06, Font06c} so any material
accreted onto the extended rotating disk would already have been more
enriched than disk-like fields.

\subsubsection{Could disk-like material originate from the thin disk?}

The previous scenarios addressing the origin of disk-like material
center upon idea that this material has been brought into M31 from
another system. However, the presence of young stars in many extended
disk fields begs the question as to whether this material could
instead have formed in the gas-rich thin disk of M31 and subsequently
been torn off and kicked out.

The G1 Clump is the only disk-like field to have stringent constraints
placed on its SFH so far. \cite{Faria07} used STARFISH modelling
\citep{HZ01} of the CMD to infer a high mean metallicity ([Fe/H] =
-0.4 dex) and a large age spread ($\sim 10$~Gyr).  Although the bulk
population is of intermediate-to-old age ($\gtrsim 6$~Gyr), roughly
10$\%$ of its stellar mass has formed within the last 2 Gyr.  As
discussed by \cite{Faria07}, these properties are entirely consistent
with those of the outer disk of M31 where trace amounts of recent star
formation are commonly observed \citep{Bellazzini03, Williams01}.  In
conjunction with its kinematic signature of disk-like rotation
\citep{Reitzel04, Ibata05}, \cite{Faria07} have suggested a connection
between the thin disk and the G1 Clump. In addition, \cite{Zucker04}
have suggested a link between the thin disk and the NE Clump as one
possible explanation of their observations. In view of the results
presented here, we explore whether this interpretation can be extended
to all of our disk-like fields.

If the disk-like material did originate in the thin disk, we require
some mechanism capable of restructuring it and moving material out to
large radii ($\sim 44~{\rm kpc}$ in the case of the NE Clump) and
scale heights. Recent work has elucidated how interactions with
sub-halos might heat, thicken and perturb stellar disks, providing a
method capable of producing such low-latitude substructure
\citep[e.g.,][]{Quinn93, Walker96, Velazquez99, Gauthier06,
Kazantzidis07}. Most recently, \cite{Kazantzidis07} have performed
hybrid cosmological plus numerical simulations of dark matter
sub-halos bombarding a MW type disk galaxy over an 8 Gyr period and
trace the impact on the host galaxy. They cull sub-halo properties
such as mass functions and orbital parameters directly from
cosmological simulations so that sub-halos are representative of their
epoch. The cumulative effect of six such bombardments, one 60$\%$ as
massive as the host's disk, is to produce several long-lived phenomena
in the disk such as a significant flare, central bar and coherent
substructures akin to tidal streams. Most damage to the thin disk is
caused by the most massive satellite. Significantly, although the thin
disk is thickened and perturbed by these tidal interactions, forming
many messy morphological features comparable to those observed in M31
and the MW, it survives intact to z = 0. \cite{Kazantzidis07} only
track the stellar particles in the original thin disk and not those of
the interacting satellites. In reality, the content of the resultant
disk will be a complex mix of stars formed in situ as well as stars
brought in by the satellite.  While more work is required to establish
the viability of this explanation, disk heating seems to explain many
of the observed properties of disk-like fields (and the extended disk
in general) at least on a qualitative level.

Many questions arise from this line of thought: Will these discrete
substructures ultimately disperse to form a thick disk in M31? Could
the bulk of the damage to M31's disk have been caused by the impact of
the GS progenitor?  It is interesting to note that current simulations
suggest the interaction between the GS progenitor and M31 began as
recently as $\sim 1~{\rm Gyr}$ ago (F07). This timescale agrees well
with the dissolution timescale of discrete substructures in the
extended disk, as estimated by \cite{Ibata05}.  However, there are
also suggestions that the GS was accreted 6-7 Gyr ago coinciding with
the apparent cessation of star formation
\citep{Brown06a}. \cite{Font07} present a comprehensive discussion of
these two competing scenarios based on the results of their numerical
simulations. The best estimates of the mass of the GS progenitor
($10^{9}$~M$_{\sun}$) place it at the low end of the sub-halo mass
range featured in \cite{Kazantzidis07}, yet it does have a highly
radial orbit which was found to enhance the disruption to the thin
disk. The simulation explained in \cite{Kazantzidis07} resulted in a
final distribution of disk stars best described by a thin~$+$~thick
disk decomposition. There is room for much speculation, but we are
unable to draw conclusions about this within the scope of this study.

\subsection{The Nature of the Composite Fields}

Our analysis has revealed that four fields -- the Brown-odisk,
GC6\_field, NGC205 Loop and Brown-spheroid -- do not fit neatly into
either stream-like or disk-like categories. Several show traits from
both. Can these populations be described as a simple combination of
disk-like and stream-like material, or are they genuinely unique
populations?

\begin{figure}[ht!]
\includegraphics[width=8.5cm]{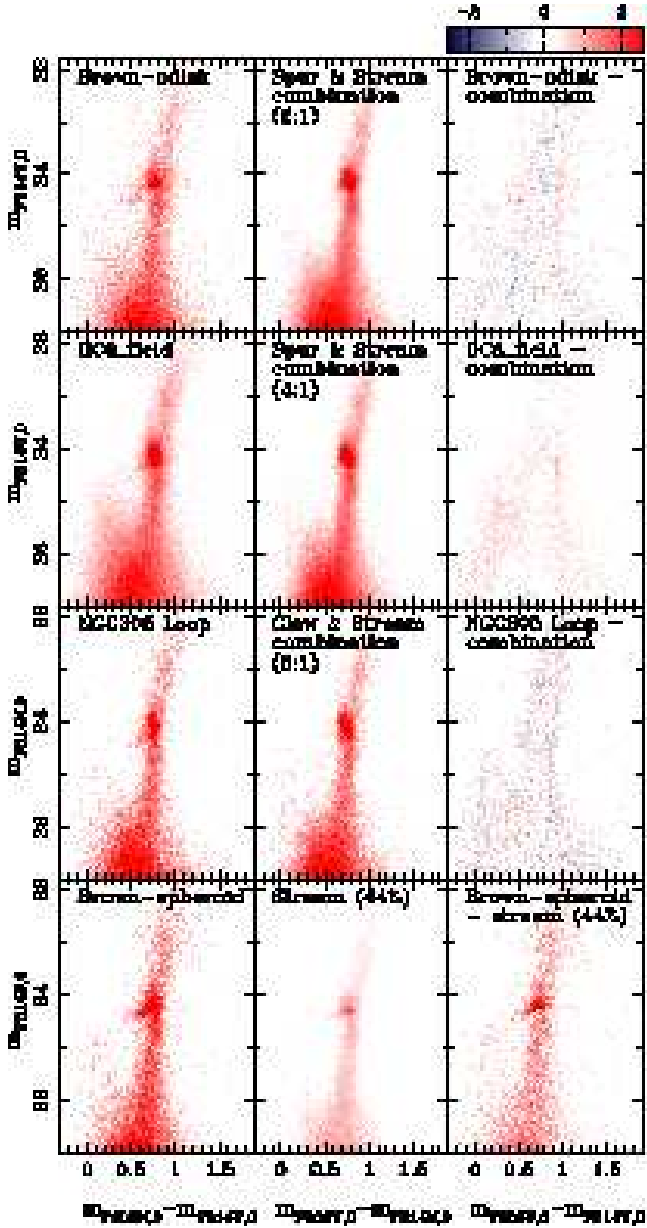}
  \caption{Examining the relative contribution of disk and stream
  material to the composite fields. The left-hand panels show the
  normalized CMDs of the composite fields while the middle panels show
  model CMDs created from differing amounts of the Spur (or Claw) and
  GS fields. The right-hand panels show the result of subtracting
  these models from the original Hess diagrams. As before, blue
  corresponds to an over-subtraction and red to an under-subtraction.}
\label{fig:composite}
\end{figure}

\subsubsection{The GC6\_field and Brown-odisk fields}
 
The CMDs of the Brown-odisk and GC6\_field exhibit many disk-like
properties, including prominent round RCs and BPs, reflecting their
location at the edge of the main M31 disk ($R_{disk}= 25.7 ~{\rm kpc}$
and 26.8 kpc respectively). However, these same fields also show
evidence of an extended HB akin to that seen in stream-like
fields. Furthermore, differencing their CMDs with the GS results in
lower amplitude over-subtractions than found in disk-like fields (see
Figure~\ref{fig:stream}) implying a minority stream-like component to
the stellar populations in these regions.  If correct, this would be
consistent with Figure~\ref{fig:fardal} which indicates that both
stream debris and thin disk stars are expected in considerable
proportions at these locations.

In order to test this further, a composite model CMD has been
constructed using the N Spur and GS fields and compared to the data in
Figure~\ref{fig:composite}. The ratio of disk to stream in the model
has been scaled by eye to produce the cleanest subtraction. Overall,
the residuals from this subtraction are of significantly lower
amplitude than those in Figure~\ref{fig:stream} implying that this
combination of fields is closer in nature to Brown-odisk and the
GC6\_field than the GS alone. In truth, these models are crude, but
they serve to illustrate that these fields are likely disk-like fields
contaminated by debris from GS. A more accurate breakdown of their
populations will be achieved via full fitting of the star formation
histories in future work.

\subsubsection{The NGC205 Loop field}

The NGC205 Loop was so named because, on projected surface density
maps, it appears as a tidal loop emanating from the dE NGC205 galaxy,
which sits a mere $\sim$ 15 kpc away. Surface brightness profiles of
NGC205 have long showed significant isophotal twisting implying tidal
stripping of its material by M31 \citep{Hodge73, Choi02}. Kinematics
of stars thought to be associated with the NGC205 Loop were presented
by \cite{McC04} who identified two kinematic signatures; a majority
component attributable to the M31 disk and a small additional
kinematically cold component ($\sigma_v \sim 10 ~{\rm km s^{-1}}$)
that they associate with the NGC205 Loop. Subsequent examination of
the same field by \cite{Ibata05} yielded no evidence for such
bimodality however and led them to the conclusion that the NGC205 Loop
was part of the same vast extended rotating disk to which the
disk-like fields belong (with $v_{lag} \sim 14~{\rm km s^{-1}}$,
$\sigma_v \sim 36 ~{\rm km s^{-1}}$).

We now turn to address what can be learnt about the origin of the
NGC205 Loop from its stellar populations.  The NGC205 Loop, despite
its stream-like CMD, shows significant residuals in the GS subtraction
which indicate a second stellar component is present at this location
(Figure~\ref{fig:stream}). Is this additional component tidal debris
from NGC205 dwarf? \cite{Butler05} have recently examined the stellar
populations in the outskirts of NGC205 and find a population
characterized by a broad RGB, prominent RC and weak HB. They estimate
a median metallicity of [Fe/H]$>-1.06$. The NGC205 Loop residuals
could be consistent with such a population. Indeed, the
under-subtracted RC in this field is slightly bluer than that seen in
the disk-like fields, suggesting the excess population is more
metal-poor than the extended disk.  On the other hand, a composite
model made from the Claw and the GS (5:1) produces a very good match
to the CMD of the NGC205 Loop in Figure~\ref{fig:composite}.  The
apparent similarity of the intrinsic NGC205 dwarf stellar population
to that of the extended disk make it difficult to draw firm
conclusions about the origin of the NGC205 Loop.  Better kinematical
constraints and more detailed modelling of NGC205's orbital trajectory
are required.


\subsubsection{The Brown-spheroid field - a glimpse of M31's
  underlying halo?}
\label{sec:sph}

The Brown-spheroid field is unlike any of the other fields presented
here. Although the GS subtraction is fairly clean, it has unveiled an
excess extended blue HB population not seen in any other subtraction.
This suggests the presence of a higher fraction of metal-poor and/or
ancient stars than in the GS material. At $R_{proj} = 11.5 ~{\rm
kpc}$, this is the innermost of our ACS pointings; if a weak smooth
halo component exists in M31, then it is in this field that we would
expect the strongest signature.

\cite{Gilbert07} have recently revisited the nature of the
Brown-spheroid field in light of their spectroscopic study of stars
along the minor axis of M31. Based on kinematics, they estimate that
$\sim$44\% of stars in this field belong to a cold component with the
same metallicity as the GS and they have shown that this fraction
agrees well with the predictions of the F07 simulation.
Figure~\ref{fig:composite} shows the result of subtracting a 44\% GS
component from the Brown-spheroid field.  The residual population is
characterized by a RC which has narrowed somewhat in luminosity, the
lack of an RGB bump and the presence of a prominent blue HB. It is
tempting to speculate that this represents the true underlying halo of
M31.

Our own Minor Axis field, which sits $\sim 8$~kpc further out than the
Brown-spheroid field, was originally intended to probe the
uncontaminated halo, however we have found here that it too shows the
signature of contamination by stream debris.  \cite{Gilbert07} trace
the kinematics in a location close to our Minor Axis field finding
that $\sim$31\% of the stars are part of a cold giant stream-like
component. It may appear surprising that this field compares so much
better to the GS than the Brown-spheroid when the fraction of stream
present is estimated to be even lower. However, it should be noted
that the Minor Axis field lies at a very similar projected radial
distance from the center of M31 as the GS field (see Table
2). Therefore, if the underlying M31 halo is spherical, the fraction
of ancient metal-poor components in these two fields should be almost
identical and thus cancel out cleanly in subtractions between the two.

\vspace{1.5 cm}

\section{Summary}
\label{sec:conc}

We have presented the largest and most detailed survey to date of
stellar populations in the outskirts of M31.  We have carried out a
homogeneous analysis of 14 deep HST/ACS pointings spanning the range
$11.5$~kpc $\lesssim $R$_{proj}\lesssim 45$~kpc. These pointings
sample well-defined substructure identified in the course of the
INT/WFC imaging survey, as well as the more diffuse extended disk.  We
have shown how stellar populations in these fields can be generally
divided into two clearly distinct categories based on their CMD
morphologies: stream-like or disk-like.  Our analysis lead us to the
following conclusions:

\begin{itemize}
\item Stream-like fields have blue RCs and extended horizontal
  branches yet are more metal enriched than disk-like fields.  They
  show no evidence of recent star formation. Our analysis reveals for
  the first time that the Western Shelf (probed by the EC1$\_$field)
  is consistent with being material torn off from the GS
  progenitor. We compare the spatial and line-of-sight distribution of
  stream-like fields with the F07 simulation of the GS progenitor
  orbit and find an excellent agreement. In this picture, the GS and
  Brown-stream fields sample the giant stream itself while the NE
  Shelf and EC1$\_$field probe the Northeastern and Western
  Shelves. The Minor Axis and Brown-spheroid fields are significantly
  contaminated by stream material that has undergone several
  pericentric passages, as previously suspected from kinematics
  \citep{Gilbert07}.
\item Disk-like fields have rounder RCs which have significant
  luminosity width and show no evidence of an extended horizontal
  branch.  All show evidence for young ($\lesssim 2$~Gyr) populations,
  and in some cases star formation as recent as 250 Myr ago. Disk-like
  material could have either an external or an internal origin.
  Recent work (e.g., \citealt{Kazantzidis07}) describes how such
  structures resembling tidal debris can be created from interactions
  between the stellar thin disk and massive satellites on close
  pericentric orbits. Given the uniform populations in these fields,
  including the ubiquitous presence of young populations, and the high
  rotation observed in the extended disk to which the disk-like fields
  belong \citep{Ibata05}, we favor this scenario over the original
  assertion that this vast extended rotating component has formed via
  accretion.
\item Several fields are identified as being `composite' in the sense
  of having both stream-like and disk-like properties.  Amongst these
  are the Brown-odisk and Brown-spheroid fields which have been the
  targets of ultra-deep HST programs aimed at determining the SFH of
  the M31 disk and halo back to the earliest times \citep{Brown06a}.
  Our findings indicate that the stellar populations in these regions
  are probably much more complicated than originally envisioned,
  making interpretation of the resulting SFHs difficult.  Another
  composite field is the NGC205 Loop. We have shown that this field
  likely contains GS contamination in addition to an underlying
  component which is consistent with being either material from NGC205
  and/or from M31's extended disk.
\end{itemize}

We have shown that much insight can be gained into the nature and
origin of substructure in M31 from the comparative analysis of CMDs
reaching a few magnitudes below the RC.  As new surveys uncover an
ever-increasing level of complexity in this system, a shallow-depth
multi-pointing approach such as this may prove very valuable. In
forthcoming work, we will present detailed SFH fits in the various
fields analyzed here as a means to confirm and better quantify their
similarities and differences. It will also be of great interest to
extend this analysis to the tidal streams recently discovered in the
far outer halo by \cite{Ibata07}.

\acknowledgements

We thank David Thilker for providing M31 HI column density values
ahead of publication and Mark Fardal for providing us with the
data-points to produce Figure~\ref{fig:fardal}.  We wish to thank Tom
Brown for discussions during the early stages of this project and
P. Stetson for providing his stand-alone DAOPHOT II code. JCR
acknowledges the award of an STFC studentship. AMNF is supported by a
Marie Curie Excellence Grant from the European Commission under
contract MCEXT-CT-2005-025869. NRT acknowledges financial support via
a STFC Senior Research Fellowship. Support for programs GO9458 and
GO10128 was provided by NASA through a grant from the Space Telescope
Science Institute, which is operated by the Association of
Universities for Research in Astronomy, Inc., under NASA contract
NAS5-26555.

\bibliographystyle{apj}
\bibliography{richapj}

\end{document}